\documentclass[]{pasj02} 
\usepackage[switch,mathlines]{lineno} 
\usepackage{xcolor}%

\jyear{2025}
\Received{2025/07/01}
\Accepted{2025/07/27}

\begin{document} 

\title{XRISM Observation of the Ophiuchus Galaxy Cluster: 
Quiescent Velocity Structure in the Dynamically Disturbed Core}

\author{
 Yutaka \textsc{Fujita},\altaffilmark{1}\altemailmark\orcid{0000-0003-0058-9719} \email{y-fujita@tmu.ac.jp} 
 Kotaro \textsc{Fukushima},\altaffilmark{2}\orcid{0000-0001-8055-7113}
 Kosuke \textsc{Sato},\altaffilmark{3}\orcid{0000-0001-5774-1633}
 Yasushi \textsc{Fukazawa}\altaffilmark{4}\orcid{0000-0002-0921-8837}
 and 
 Marie \textsc{Kondo}\altaffilmark{5}\orcid{0009-0005-5685-1562}
}
\altaffiltext{1}{Department of Physics, Graduate School of Science,
Tokyo Metropolitan University,
1-1 Minami-Osawa, Hachioji-shi, Tokyo 192-0397}
\altaffiltext{2}{Institute of Space and Astronautical Science, JAXA, 3-1-1 Yoshinodai, Chuo-ku, Sagamihara, Kanagawa 252-5210}
\altaffiltext{3}{Department of Astrophysics and Atmospheric Sciences,
Kyoto Sangyo University, Kamigamo-motoyama, Kita-ku, Kyoto 603-8555}
\altaffiltext{4}{Department of Physical Science, Hiroshima University, 1-3-1 Kagamiyama, Higashi-Hiroshima, Hiroshima 739-8526}
\altaffiltext{5}{Graduate School of Science and Engineering, Saitama University, 255 Shimo-Ohkubo, Sakura, Saitama 338-8570}




\maketitle

\begin{abstract}
We present the high-resolution X-ray spectroscopic observations of the Ophiuchus galaxy cluster core using the XRISM satellite. Despite previous observations revealing multiple cold fronts and dynamical disturbances in the cluster core, our XRISM observations show low gas velocity dispersions of $\sigma_v = 115\pm 7$ km s$^{-1}$ in the inner region ($\lesssim 25$ kpc) and $\sigma_v = 186\pm 9$ km s$^{-1}$ in the outer region ($\sim$ 25--50 kpc). The gas temperatures are $kT = 5.8\pm 0.2$ keV and $8.4\pm 0.2$ keV for the inner and outer regions, respectively, with metal abundances of $Z = 0.75\pm 0.03\: Z_\odot$ (inner) and $0.44\pm 0.02\: Z_\odot$ (outer). The measured velocity dispersions correspond to nonthermal pressure fractions of only $1.4\pm 0.2\:$\% (inner) and $2.5\pm 0.2\:$\% (outer), indicating highly subsonic turbulence. Our analysis of the bulk gas motion indicates that the gas in the inner region is nearly at rest relative to the central galaxy ($|v_{\rm bulk}|=8\pm 7$~km s$^{-1}$), while the outer region exhibits a moderate motion of $|v_{\rm bulk}|=104\pm 7$~km s$^{-1}$. Assuming the velocity dispersion arises from turbulent motions, the turbulent heating rate is $\sim$~40\% of the radiative cooling rate, although there is some uncertainty. This suggests that the heating and cooling of the gas are not currently balanced. The activity of the central active galactic nucleus (AGN) has apparently weakened. The sloshing motion that created the cold fronts may now be approaching a turning point at which the velocity is minimum. Alternatively, the  central galaxy and the associated hot gas could be moving nearly parallel to the plane of the sky.
\end{abstract}


\section{Introduction}

Galaxy clusters are the largest gravitationally bound structures in the universe, representing the endpoints of cosmic structure formation \citep{2012ARA&A..50..353K}. These massive systems, with masses ranging from $10^{14}$ to $10^{15}\: M_{\odot}$, are dominated by dark matter but contain significant amounts of hot, X-ray emitting gas known as the intracluster medium (ICM). The ICM is heated to temperatures of 10$^7$-10$^8$ K through gravitational processes during cluster assembly \citep{1988xrec.book.....S}.

The physics of the ICM is governed by a complex interplay of gravitational heating, radiative cooling, and various feedback mechanisms. In the central regions of many clusters, the ICM exhibits a ``cool core'' structure characterized by a sharp increase in gas density and decrease in temperature toward the cluster center. These conditions lead to radiative cooling times that are much shorter than the Hubble time, theoretically resulting in massive cooling flows with rates of hundreds to thousands of solar masses per year \citep{1994ARA&A..32..277F}.

However, observations consistently reveal a fundamental discrepancy known as the ``cooling flow problem.'' Despite the short cooling times in cluster cores, line emissions from cooling gas are not observed \citep{1997ApJ...481..660I,2001A&A...365L..99K,2001A&A...365L.104P,2001A&A...365L..87T}. This suggests that some form of heating mechanism must be operating to prevent or significantly reduce the cooling flow.

Active galactic nuclei (AGN) feedback has emerged as the leading solution to the cooling flow problem. The supermassive black holes at the centers of brightest cluster galaxies (BCGs) can inject enormous amounts of energy into the surrounding ICM through relativistic jets and outflows \citep{2000A&A...356..788C,2007ARA&A..45..117M,2012ARA&A..50..455F}. These AGN-driven processes create X-ray cavities, inflate radio bubbles, and drive shock waves and turbulence throughout the cluster core. The energy deposited by AGN activity appears to be sufficient to offset radiative cooling losses in many systems.

However, the detailed mechanisms by which AGN energy is transferred to the ICM remain poorly understood. The energy must somehow be converted from the highly collimated, relativistic jets into thermal energy that can heat the diffuse gas. Proposed mechanisms include the buoyant rise and mixing of relativistic plasma, cosmic rays, and the dissipation of turbulent motions \citep{2001ApJ...554..261C,2017MNRAS.466L..39H,2013MNRAS.428..599F,2014Natur.515...85Z}.

Direct measurements of gas velocities in the ICM offer a unique window into these heating processes. High-resolution X-ray spectroscopy can detect both bulk motions (through Doppler shifts of emission lines) and turbulent velocities (through line broadening). Until recently, such measurements were extremely challenging due to the limited spectral resolution of CCD detectors on missions like Chandra, XMM-Newton, and Suzaku.

The situation changed dramatically with the launch of the Hitomi satellite in 2016, which carried a microcalorimeter capable of achieving spectral resolution of $\sim$~5 eV \citep{2016Natur.535..117H}. Although the satellite was short-lived, Hitomi's observations of the cool core of the Perseus cluster provided the direct measurement of ICM turbulence, revealing surprisingly low velocity dispersions of $\sim$~160 km s$^{-1}$ and nonthermal pressure support of only $\sim$~4\% \citep{2016Natur.535..117H,2018PASJ...70....9H}. The observational results of the XRISM satellite \citep{2025PASJ..tmp...28T}, which is the recovery mission of Hitomi, have recently been reported. The velocity dispersions of the cool cores of the Centaurus, Abell~2029, and Hydra~A clusters are only $\sim$~120--170 km s$^{-1}$ \citep{2025Natur.638..365X,2025ApJ...982L...5X,2025arXiv250501494R}. Even for the Coma cluster, which is a merging cluster, the dispersion is $\sim 200$~km s$^{-1}$ \citep{2025ApJ...985L..20X}.

The Ophiuchus cluster is a nearby galaxy cluster ($z=0.0296$; \cite{2015A&A...583A.124D}) that possesses a cool core despite being a high-temperature cluster ($kT\sim 9$~keV; \cite{2008PASJ...60.1133F}). Previous Chandra observations have detected multiple cold fronts and dynamical disturbances within the cool core, suggesting some form of gas motion (\cite{2016MNRAS.460.2752W}; see also figure~\ref{fig:image}). Radio observations have also reported evidence of past AGN activity of unprecedented scale ($\sim 5\times 10^{61}$~erg) at the cluster's center \citep{2020ApJ...891....1G}. In this letter, we present the results of our XRISM observations of the Ophiuchus cluster. Our focus is on the velocity structure of the cool core. Other topics, such as metal abundance patterns, will be explored in future work. The cosmological parameters we adopted are $\Omega_0=0.3$, $\Lambda=0.7$, and $h=0.7$. Thus, $1'$ corresponds to 35.6 kpc at the redshift of the Ophiuchus cluster ($z=0.0296$). The BCG, or central galaxy of the cluster, is 2MASX J17122774-2322108, with a redshift of $z=0.0295$ \citep{2015A&A...583A.124D}. We use the proto-solar abundance table from \citet{2009LanB...4B..712L}. The quoted errors represent the $1\:\sigma$ statistical uncertainties, unless otherwise mentioned.

\section{Observations and Data Reduction}

\begin{figure}
 \begin{center}
  \includegraphics[width=8cm]{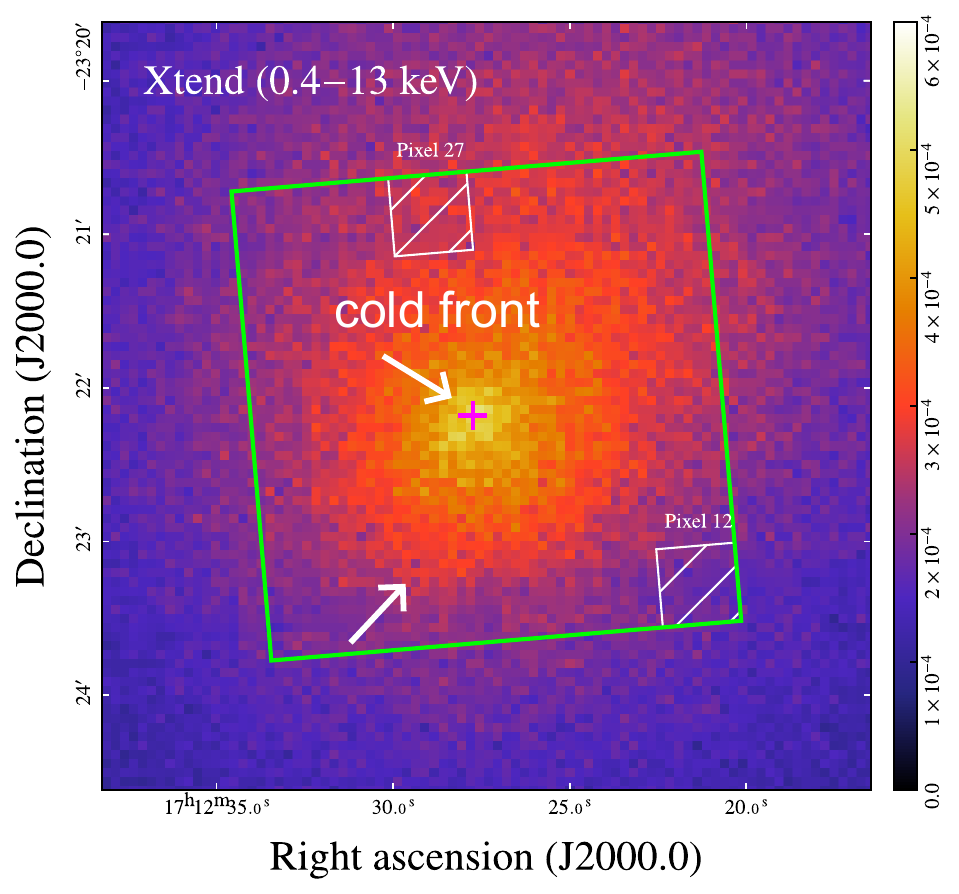} 
 \end{center}
\caption{XRISM Xtend image (0.4--13~keV) of the Ophiuchus cluster overlaid with the Resolve FOV (green box). The magenta cross indicates the X-ray peak. White arrows indicate cold fronts. The positions of pixels 12 and 27 are shown.
} 
\label{fig:image}
\end{figure}

XRISM observed the central region of the Ophiuchus cluster from March 31 to April 6, 2025 (obsid 201006010), with the aim-point positioned at $\alpha=258.115^\circ$, $\delta=-23.3687^\circ$ at the cluster's X-ray center (figure~\ref{fig:image}). In this paper, we focus on data from the Resolve instrument, a microcalorimeter array that covers a $3.1'\times 3.1'$ field of view (FOV) with a $6\times 6$ pixel configuration \citep{2022SPIE12181E..1SI}. Each pixel produces a spectrum of incident X-rays with a resolution of 4.5 eV FWHM. The instrument's energy band spans 1.7--12 keV. At low energies, it is limited by the attenuation of the dewar gate valve, which is currently closed. The Resolve data were processed with the latest version 3 software (PROCVER=03.00.013.010) and analyzed using XRISM FTOOLS wrapped in HEASoft version 6.34 and calibration database (CALDB) version 11. Following the standard screening procedures outlined in the XRISM Data Reduction Guide \footnote{https://heasarc.gsfc.nasa.gov/docs/xrism/analysis/index.html}, the observation yielded a cleaned exposure time of 217 ksec. 

Each spectrum was obtained by integrating the high-resolution (Hp) grade events across all pixels in the Resolve FOV. We generated spectral redistribution matrix files (RMFs) using the \texttt{rslmkrmf} task with ``L'' size option. The ARF files calculate the effective area by incorporating detector efficiencies and the response of the X-ray Mirror Assembly (XMA), including point spread function (PSF) effects. These files were generated by \texttt{xaarfgen} task using the X-ray image from the Chandra satellite as the input brightness distribution. The barycentric correction of $28\rm\: km\: s^{-1}$ is applied to X-ray velocities in this paper.

\begin{figure}
 \begin{center}
  \includegraphics[width=8cm]{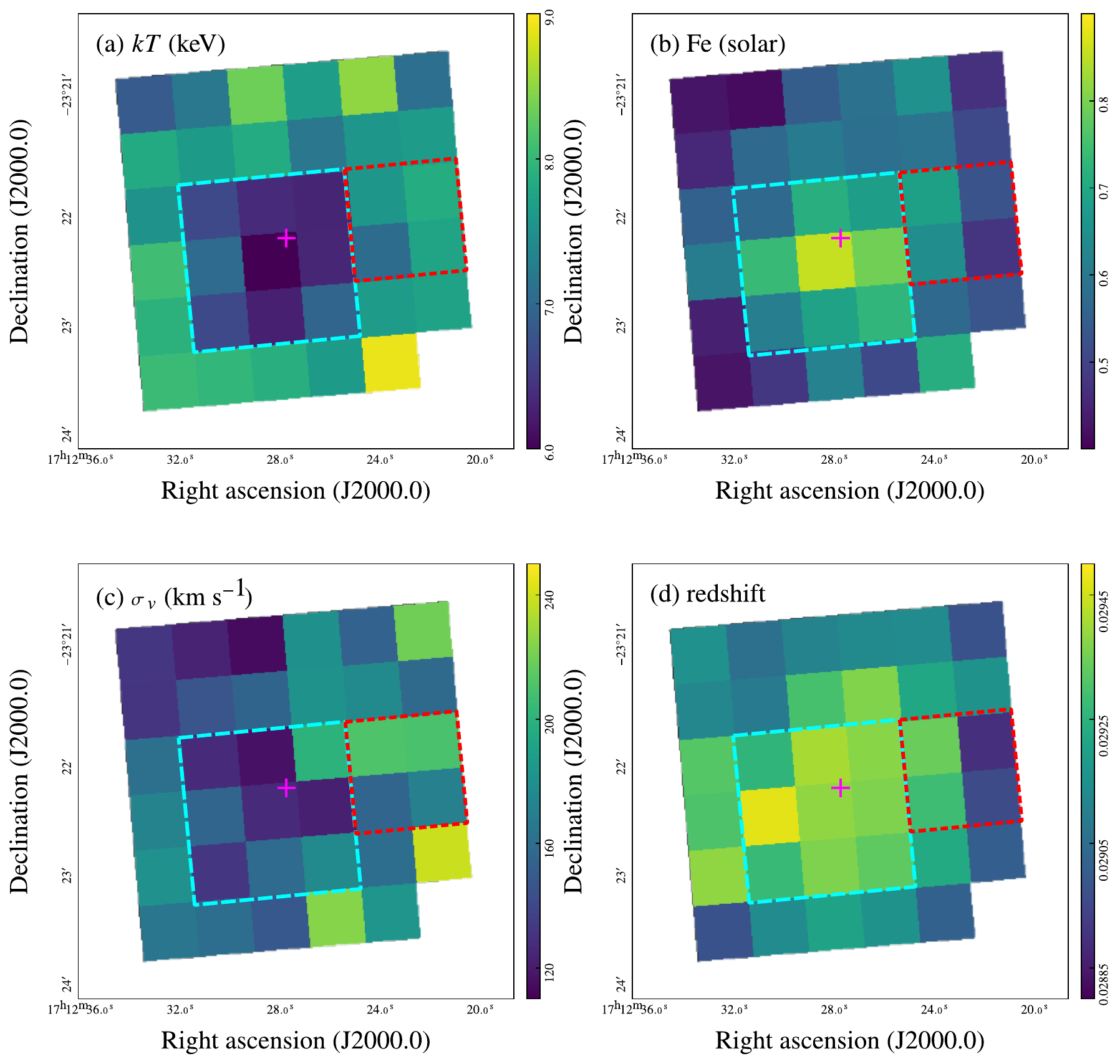} 
 \end{center}
\caption{Maps of (a) temperature, (b) metal abundance, (c) (turbulent) velocity dispersion, and (d) redshift of the ICM obtained with Resolve. The magenta cross shows the X-ray peak, and the cyan dashed square shows the boundary between the inner and outer regions. The red dotted square indicates the area where an peculiar iron line feature was discovered.
} 
\label{fig:map}
\end{figure}

\section{Spectral Modeling and Analysis}

First, we created maps depicting the temperature ($kT$), metal abundance ($Z$), velocity dispersion ($\sigma_v$), and redshift of the ICM ($z$) in order to understand its overall properties. We obtained these maps through a pixel-by-pixel spectral analysis of the Resolve data using a thermal model (bapec). Figure \ref{fig:map} shows the results. The temperature is lower and the abundance is higher in the central nine pixels, which are shifted slightly southeast of the X-ray peak (cyan dashed square). We refer to the region inside the square as the ``inner region'' and the region outside the square as the ``outer region''. We note that pixel 27 has been identified to show irregular variation of the energy scale during the observation, which is hard to track using the current gain-monitoring procedure. Thus, we exclude this pixel from the subsequent analysis. We also excluded calibration pixel 12 (figure~\ref{fig:image}).

We performed a detailed spectral analysis for the two regions. Because the PSF of XRISM is relatively large ($\sim 1.3'$ for the half-power diameter), a significant number of photons emitted from outside a given region may be detected by the telescope as being within the region, and vice versa. We accounted for the spatial-spectral mixing (SSM) effect by fitting the spectra of the two regions simultaneously with plasma models and appropriate weights. Although we took the charged-particle-induced non-X-ray background (NXB) into account, we confirmed that it had little impact on the results. We fit the Resolve spectra with a thermal model (\texttt{bapec}) absorbed by photoelectric absorption (\texttt{phabs}) using Xspec v12.14.1 \citep{1996ASPC..101...17A} and employed C-statistics \citep{1979ApJ...228..939C}. The Galactic hydrogen column was fixed to the value
of $N_{\rm H}=1.9\times 10^{21}\rm\: cm^{-2}$ \citep{2016A&A...594A.116H}.
This study focuses on four key parameters: temperature ($kT$), abundance ($Z$), velocity dispersion ($\sigma_v$), and bulk velocity ($v_{\rm bulk}$). The latter is estimated from redshift ($z$) and indicates the systematic velocity relative to the BCG.

\begin{figure*}[ht!]
 \begin{center}
  \includegraphics[width=\linewidth]{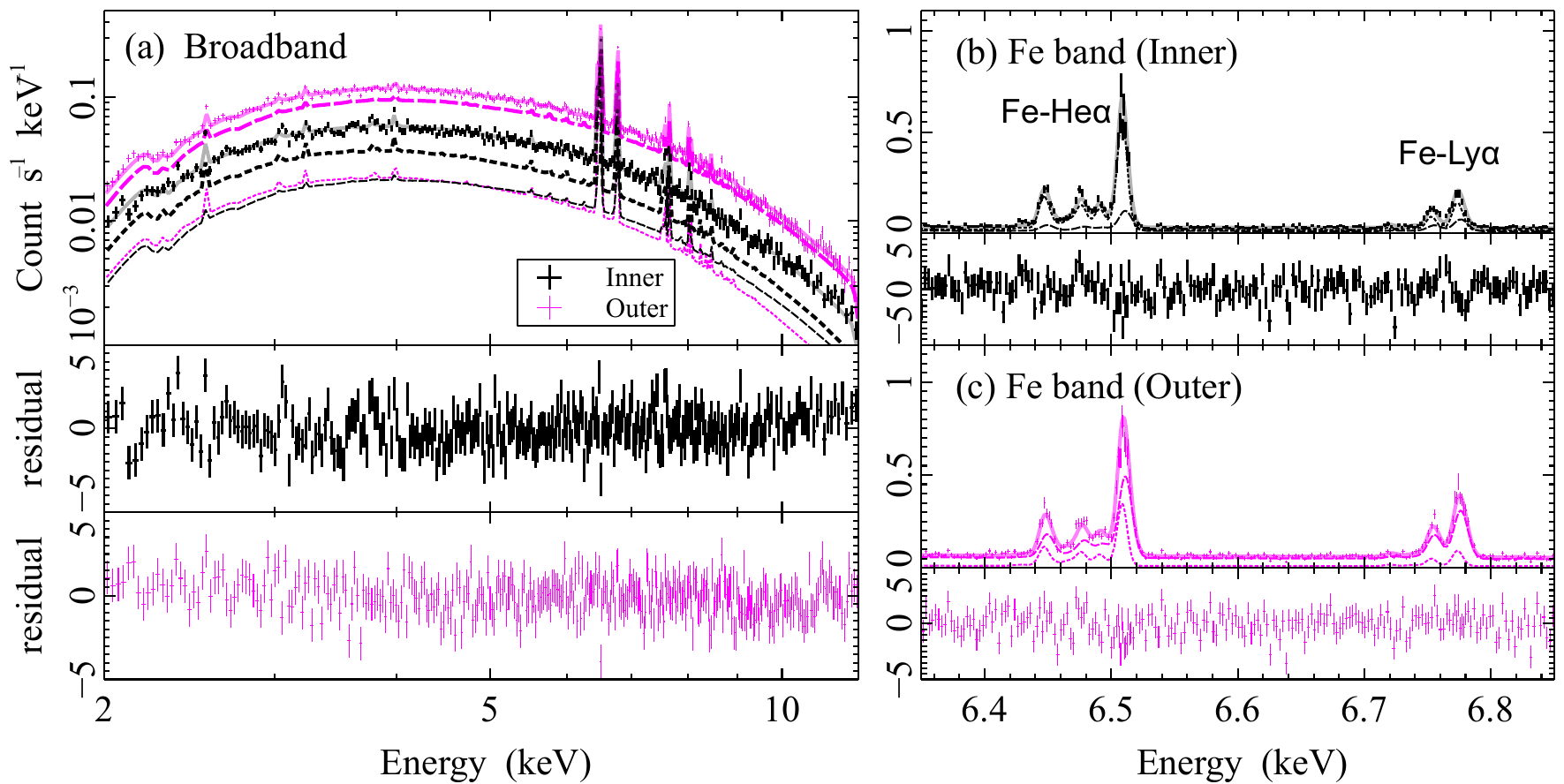} 
 \end{center}
\caption{Resolve spectra and the best-fitting models, including the SSM effect. (a) Broadband spectra. The black lines show the spectrum of the inner region. From top to bottom, they represent the total spectrum, the inner region's contribution, and the outer region's contribution. The magenta lines show the spectrum of the outer region. From top to bottom, they represent the total spectrum, the outer region's contribution, and the inner region's contribution. (b) and (c) Zoom-in spectra around Fe lines.}
\label{fig:spec}
\end{figure*}

\section{Results}

Figure~\ref{fig:spec}a shows the results of fitting the spectra of the two regions simultaneously in the 2--12 keV range, accounting for the SSM effects. Table~\ref{tab:results} summarizes the best-fitting parameters. The results do not change within the statistical errors, even when using the 5--12 keV range, because the most prominent lines are Fe lines at $\sim 6.4$--6.8 keV (figures~\ref{fig:spec}b and c).
As indicated by previous observations \citep{2008PASJ...60.1133F, 2010MNRAS.405.1624M, 2016MNRAS.460.2752W}, the temperature is lower and the abundance is higher in the inner region compared to the outer region, which is typical for a cool core cluster. The stronger Fe Ly$\alpha$ lines observed in the outer region (figure~\ref{fig:spec}c) reflect a higher temperature than the inner region (figure~\ref{fig:spec}b). The temperature gradient indicates substantial cooling in the central region. The abundance gradient probably reflects the enrichment history of the cluster core, with metals produced by stellar nucleosynthesis in the BCG.

The most significant and surprising result from our XRISM observations is the remarkably low level of gas velocity dispersion in both regions of the Ophiuchus cluster core. The inner region exhibits a velocity dispersion of $\sigma_v = 115 \pm 7$ km s$^{-1}$. This is similar to that of the Centaurus cluster ($\sigma_v \lesssim 120\rm\: km\: s^{-1}$), in which the central AGN activity is weak \citep{2025Natur.638..365X}. 
For the temperature of 5.8 keV, the sound speed is $c_s=1240\rm\: km\: s^{-1}$. Assuming the velocity dispersion determined by Resolve ($\sigma_v = 115$ km s$^{-1}$) is entirely due to isotropic turbulence, our measurement implies a turbulent Mach number of $\mathcal{M}_{\rm 3D}=\sqrt{3}\: \sigma_v/c_s=0.16$, indicating that the motions are highly subsonic. The resulting nonthermal (NT) pressure fraction is
\begin{equation}
 \frac{P_{\rm NT}}{P_{\rm tot}} = \frac{\mathcal{M}_{\rm 3D}^2}{\mathcal{M}_{\rm 3D}^2 + 3/\gamma}=1.4\pm 0.2\%\:,
\end{equation}
where $\gamma=5/3$ is the adiabatic index (e.g. \cite{2019A&A...621A..40E}). This value is smaller than those for the central region of the Centaurus cluster ($3.3\pm 0.6\:\%$; \cite{2025Natur.638..365X}) and Abell 2029 ($2.6\pm 0.3\:\%$; \cite{2025ApJ...982L...5X}); the latter is one of the most dynamically relaxed clusters. It is also significantly smaller than the values for clusters with powerful AGNs, such as Perseus ($3.9\pm 0.8\:\%$; \cite{2016Natur.535..117H}) and Hydra A ($4.5\pm 0.5\:\%$; \cite{2025arXiv250501494R}).

For the outer region ($\sigma_v = 186$ km s$^{-1}$ and $kT = 8.4$ keV; see table~\ref{tab:results}), the values are $c_s=1500\rm\: km\: s^{-1}$, $\mathcal{M}_{\rm 3D}=0.22$, and $P_{\rm NT}/P_{\rm tot}=2.5\pm 0.2\:\%$. The value of $P_{\rm NT}/P_{\rm tot}$ is still small and comparable to that of Abell 2029.
Interestingly, \citet{2016MNRAS.460.2752W} reported that $\sigma_v \lesssim 100\rm\: km\: s^{-1}$ based on surface brightness fluctuation analysis. However, they studied the region of $r=1.5$--5 arcmin, most of which is outside the Resolve FOV.

Bulk velocity measurements reveal the relative velocity between the ICM and the BCG. The inner region exhibits an extremely small velocity of $v_{\rm bulk} = +8 \pm 7$ km s$^{-1}$. This indicates that the core gas is essentially at rest with respect to the BCG. The Mach number is only $\mathcal{M}_{\rm bulk}=v_{\rm bulk}/c_s=0.007$.
For the outer region, the bulk velocity is $v_{\rm bulk} = -104 \pm 7\rm\: km\: s^{-1}$ and the Mach number is $M_{\rm bulk} = 0.07$. These bulk velocities are consistent with previous XMM-Newton observations using the EPIC-pn detector, though the velocity uncertainties are significant \citep{2023MNRAS.522.2325G}. 

Considering the steep temperature and abundance gradients at the center of the Ophiuchus cluster \citep{2016MNRAS.460.2752W}, we conducted a two-component spectral analysis of the inner region, in which $kT$, $\sigma_v$, and $v_{\rm bulk}$ are independent. The abundance $Z$ is common because that of the lower temperature component cannot be determined. The results are $kT=2.1_{-0.2}^{+0.4}$~keV, $\sigma_v=111\pm 45\rm\: km\: s^{-1}$, and $v_{\rm bulk}=-41\pm 48\rm\: km\: s^{-1}$ for the lower-temperature component, while $kT=6.5_{-0.2}^{+0.3}$~keV, $\sigma_v=112\pm 8\rm\: km\: s^{-1}$, and $v_{\rm bulk}=+12\pm 8\rm\: km\: s^{-1}$ for the higher-temperature component. The abundance is $Z=0.84\pm 0.03\: Z_\odot$. The values of $\sigma_v$ and $v_{\rm bulk}$ for both components are consistent with those of the inner region in table~\ref{tab:results}. Therefore, it is unlikely that the steep gradients affect the small values of $\sigma_v$ and $|v_{\rm bulk}|$.

Following previous studies \citep{2010SSRv..157..193C,2013MNRAS.435.3111Z,2018PASJ...70...10H}, we calculated the radial profile of the optical depth of the $w$ resonance line within the He-like Fe $\alpha$ line complex (see figure~\ref{fig:Fe}). The line is expected to be most influenced by resonant scattering. In our calculations, we used the ICM density, temperature, and abundance profiles obtained by \citet{2016MNRAS.460.2752W} and assumed $\sigma_v=0$. We found that the line is optically thick only in the central few kiloparsecs. Since the regions we studied are much larger, we expect the influence of resonant scattering to be minimal. In fact, we fit the Resolve spectrum of the inner region by replacing the $w$ resonance line with a Gaussian (\texttt{zgauss}) to include the effect of resonant scattering. We confirmed that the results did not change within the statistical errors.

\begin{table}
  \tbl{Best-fit spectral parameters for the Ophiuchus cluster core.}{%
  \begin{tabular}{ccc}
      \hline
Parameter & Inner Region & Outer Region \\
& ($r \lesssim 25$ kpc) & ($25 \lesssim r \lesssim 50$ kpc) \\
\hline
$kT$ (keV) & $5.8 \pm 0.2$ & $8.4 \pm 0.2$ \\
$Z$ ($Z_{\odot}$) & $0.75 \pm 0.03$ & $0.44 \pm 0.02$ \\
$\sigma_v$ (km s$^{-1}$) & $115 \pm 7$ & $186 \pm 9$ \\
$P_{\rm NT}/P_{\rm tot}$ (\%) & $1.4\pm 0.2$ & $2.5\pm 0.2$ \\
$v_{\rm bulk}$ (km s$^{-1}$) & $+8 \pm 7$ & $-104 \pm 7$ \\
      \hline
    \end{tabular}}\label{tab:results}
\begin{tabnote}
C-stat is 30828.60 with 32562 degrees of freedom.
\end{tabnote}
\end{table}

\section{Discussion}

\subsection{Velocity Structure}

Our XRISM Resolve observations of the Ophiuchus cluster core revealed a quiescent velocity structure. While turbulence could be created by X-ray cavities associated with AGN activity, the extremely low level of turbulence in the inner region ($P_{\rm NT}/P_{\rm tot}=1.4\:\%$; table~\ref{tab:results}) suggests that the central AGN has little kinematic impact on the surrounding ICM. We note that the level of turbulence ($1.4\:\%$) is nearly the lower limit predicted by TNG-cluster simulations for Perseus-like clusters ($\gtrsim 2\:\%$; see figure~4 in \cite{2024A&A...686A.200T}).

Turbulence dissipation is expected to heat the gas \citep{2014Natur.515...85Z}. The turbulent heating rate of gas with a mass density $\rho$ is $Q_{\rm turb} \sim 5\: \rho \sigma_v^3/l_t$, where $l_t$ is the length scale \citep{2014Natur.515...85Z}. The radiative cooling rate of the gas is calculated from the gas density and temperature $T$: $Q_{\rm cool}=n_e n_i \Lambda_n(T)$, where $n_e$ and $n_i$are the electron and ion number densities, respectively, and $\Lambda_n(T)$ is the normalized cooling function \citep{1993ApJS...88..253S}.
For the inner region, $\sigma_v\sim 115\rm\: km\: s^{-1}$, $l_t\sim 25$~kpc (the size of the inner region), the temperature $kT\sim 5.8$~keV and the metal abundance $Z\sim 0.75$~solar, while $\rho$, $n_e$ and $n_i$ are obtained from previous Chandra observations ($n_e\sim 0.03\rm\: cm^{-3}$; \cite{2016MNRAS.460.2752W}). 
From these values, we obtain $Q_{\rm turb}/Q_{\rm cool} \sim 0.4$, though $Q_{\rm turb}$ depends on the assumption of $l_t$. This value suggests that turbulent heating could not be ignored in order to counteract radiative cooling. 
The small value of $Q_{\rm turb}/Q_{\rm cool}<1$ is consistent with the absence of observable X-ray cavities in the core and the presence of only weak, point-like radio emission without lobes or jets \citep{2016MNRAS.460.2752W}. Chandra observations have revealed steep temperature and abundance gradients in the cluster core. These gradients suggest that radiative cooling is ongoing and largely unaffected by AGN heating \citep{2010MNRAS.405.1624M,2016MNRAS.460.2752W}. In other words, new powerful AGN activity has not yet been triggered, which suggests that cooling and heating may currently be imbalanced. Another possibility is that other heating mechanisms are at work, such as cosmic-ray streaming \citep{1991ApJ...377..392L,2008MNRAS.384..251G,2013MNRAS.432.1434F,2017MNRAS.467.1478J,2017ApJ...844...13R} or thermal conduction \citep{1979PThPh..62.1253T,2001ApJ...562L.129N}.

Cold fronts are contact discontinuities in the ICM and are often considered a sign of bulk motion of the ICM \citep{2000ApJ...541..542M,2001ApJ...562L.153M,2001ApJ...551..160V}. In the case of the Centaurus cluster, which has multiple cold fronts in its core, oscillating gas motion, or ``sloshing,'' has caused relative motion ($|v_{\rm bulk}|\sim 130$--$310\rm\: km\: s^{-1}$) between the ICM and the BCG \citep{2025Natur.638..365X}.
In contrast, despite the presence of multiple cold fronts, the absolute bulk velocity of the ICM in the cool core of the Ophiucchus cluster is very small
 ($|v_{\rm bulk}|\lesssim 100\rm\: km\: s^{-1}$; table~\ref{tab:results}).
\citet{2015A&A...583A.124D} indicated that the line-of-sight velocity of the BCG is consistent with the mean cluster velocity within the errors, with a difference of $\Delta v = 47\pm 97\rm\: km\: s^{-1}$. These facts could be explained if the inner core, including the BCG, were at rest at the bottom of the cluster's gravitational potential well. On the other hand, the sloshing motion of the ICM, induced by past cluster mergers, produced the cold fronts \citep{2004ApJ...612L...9F} and weakened the AGN activity by shifting the gas accretion center \citep{2010MNRAS.405.1624M,2012MNRAS.421.3409H,2016MNRAS.460.2752W}. The sloshing motion may now be approaching a turning point at which the velocity is minimum.
The velocity dispersion ($\sigma_v$) and the absolute bulk velocity ($|v_{\rm bulk}|$) are both larger in the outer region than in the inner region (table~\ref{tab:results}). This may be a remnant of the interaction between the core and the surrounding ICM.
Another possibility is that the BCG and the associated hot gas are moving almost exactly along the plane of the sky.

\begin{figure}
 \begin{center}
  \includegraphics[width=8cm]{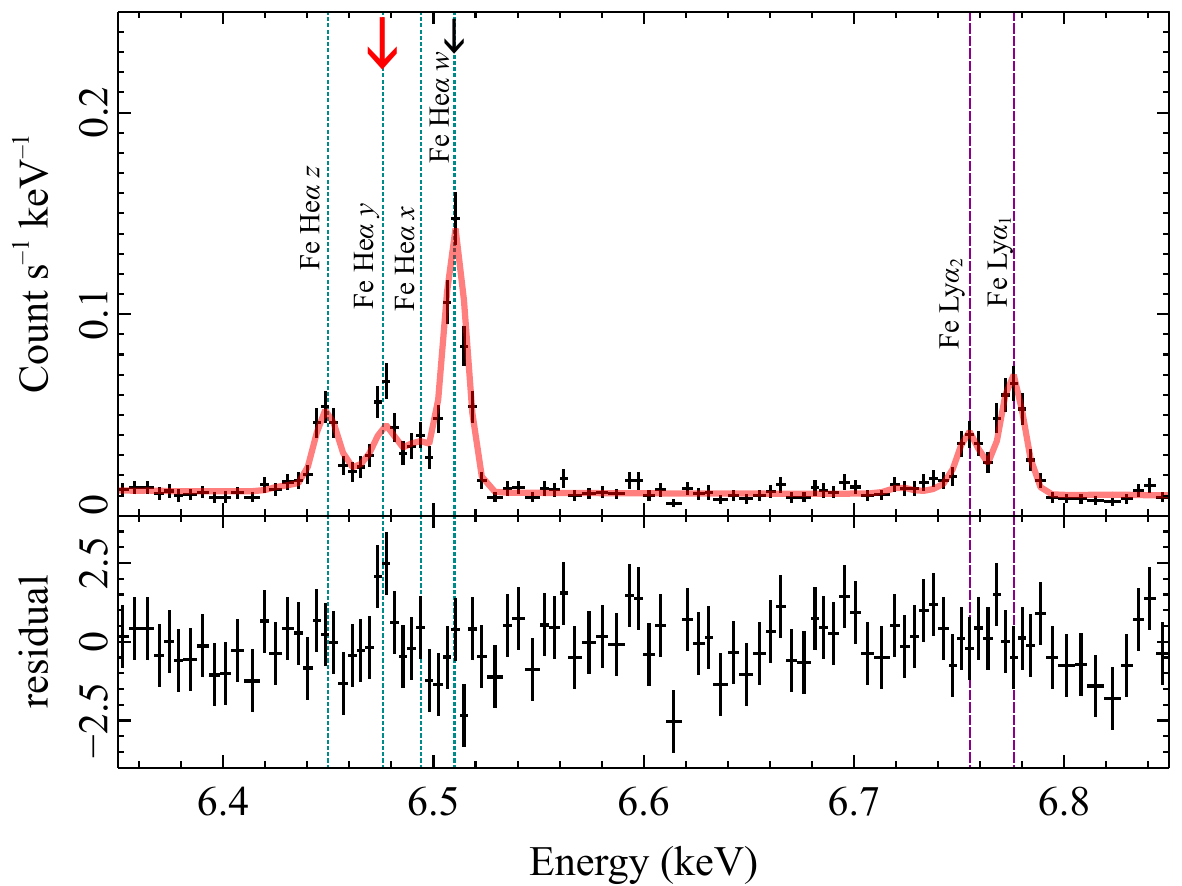} 
 \end{center}
\caption{The spectrum for the west of the cluster center
(the region shown by the red dotted square in figure~\ref{fig:map}). Only the Fe-He$\alpha$ and Fe-Ly$\alpha$ line complexes are displayed.The red arrow indicates the $y$ intercombination line, and the black arrow indicates the $w$ resonance line within the He-like Fe K$\alpha$ line complex.} 
\label{fig:Fe}
\end{figure}

\subsection{Peculiiar Iron Line Features}

XRISM has revealed that ICM spectra often exhibit features that cannot be explained by known physics, such as resonance scattering. For example, the observed flux of the $y$ intercombination line within the He-like Fe K$\alpha$ line complex for Abell 2029 is considerably higher than predicted by the standard collisional ionization equilibrium model \citep{2025ApJ...982L...5X}. An apparent excess of the Fe Ly$\alpha_2$ line flux over the model has been observed in the Coma cluster \citep{2025ApJ...985L..20X}.

Motivated by the previous studies, we investigated the ICM spectra from $2\times 2$ pixels across the Resolve FOV and found an abnormal feature in the region indicated by the red dotted square in figure~\ref{fig:map}. Figure~\ref{fig:Fe} shows an excess of the $y$ intercombination line, as found in Abell 2029. However, this specific region in figure~\ref{fig:map} does not appear particularly unusual, and the origin of the excess is unknown at this time. Such an anomaly will be examined in greater detail using several XRISM observations of other galaxy clusters. Furthermore, laboratory plasma spectroscopy using electron beam ion traps, for instance, will expand our fundamental understanding of atomic physics and emission processes. These findings highlight the importance of collaborative efforts with the atomic physics communities in the XRISM era.

\section{Conclusion}

Our high-resolution X-ray spectroscopic observations of the Ophiuchus cluster core, conducted with the XRISM satellite, have provided valuable insights into the dynamical state of this system. Despite the presence of previously identified cold fronts and dynamical disturbances, our measurements show that the ICM is remarkably quiescent in the core.

The key findings of this study include the detection of low gas velocity dispersions of $\sigma_v = 115\pm 7$ km s$^{-1}$ in the inner region ($\lesssim 25$ kpc) and $186\pm 9$ km s$^{-1}$ in the outer region ($\sim 25$--50 kpc) of the core. These measurements correspond to exceptionally low nonthermal pressure fractions of $1.4\pm 0.2\:\%$ and $2.5\pm 0.2\:\%$, respectively, indicating that the turbulent motions are highly subsonic. The gas in the inner region is nearly at rest relative to the central galaxy ($|v_{\rm bulk}|=8\pm 7$ km s$^{-1}$), while the outer region exhibits moderate motion ($|v_{\rm bulk}|=104\pm 7$ km s$^{-1}$). The gas temperatures are $kT = 5.8\pm 0.2$ keV in the inner region and $8.4\pm 0.2$ keV in the outer region, and the metal abundances are $Z = 0.75\pm 0.03\: Z_\odot$ in the inner region and $0.44\pm 0.02\: Z_\odot$ in the outer region. These gradients indicate that this is a cool-core cluster. 

Using the measured velocity dispersion, we estimate that the turbulent heating rate is only $\sim$40\% of the radiative cooling rate ($Q_{\rm turb}/Q_{\rm cool}\sim 0.4$), although there is some uncertainty. This energy deficit suggests that turbulent heating is currently unable to maintain thermal equilibrium in the cool core, and is consistent with the absence of observable X-ray cavities and the presence of only weak, point-like radio emission without extended lobes or jets. The current thermal imbalance suggests that the cluster may be in a transitional phase in which AGN activity has weakened and cooling has dominated. This indicates that clusters may undergo cyclical phases of heating and cooling imbalance.

Unlike the Centaurus cluster, where cold fronts are associated with significant sloshing gas motion ($|v_{\rm bulk}|\sim 130$--$310$ km s$^{-1}$), the Ophiuchus cluster exhibits cold fronts despite very low bulk velocities. This suggests that the sloshing motion responsible for creating these cold fronts may be approaching a turning point at which the velocity reaches its minimum. This sloshing motion may also have contributed to the weakening of AGN activity by displacing the gas accretion center.
Alternatively, the BCG and the associated hot gas could be moving nearly parallel to the plane of the sky.

Our spectroscopic analysis revealed intriguing anomalous features in the Fe line complex. We identified an excess of the $y$ intercombination line within the He-like Fe K$\alpha$ complex in a specific region. This phenomenon is similar to that observed in Abell 2029; however, the origin of this excess remains unclear.

\begin{ack}
We would like to thank the anonymous referee for their useful comments. The authors would also like to thank Kazunori Suda for his estimation of the resonant scattering and Yoshiaki Kanemaru for generating the Xtend image. The authors thank Norbert Werner and John ZuHone for their helpful
discussions. This work was supported JSPS KAKENHI grant Nos. 22H00158,
23H04899, 25H00672 (Y.~Fujita),
\end{ack}


\end{document}